\documentclass[a4paper]{jpconf}
\usepackage{graphicx}
\usepackage{physics}
\usepackage{cite}
\usepackage[export]{adjustbox}

\begin{document}
\title{Beyond-Mean-Field with an Effective Hamiltonian Mapped from an Energy Density Functional}

\author{J.~Ljungberg, J.~Boström, B.~G.~Carlsson, A.~Idini and J.~Rotureau}

\address{Mathematical Physics, LTH, Lund University, S-22100 Lund, Sweden}

\ead{jimmy.ljungberg@matfys.lth.se}

\begin{abstract}
A method for beyond-mean-field calculations based on an energy density functional is described. The main idea is to map the energy surface for the nuclear quadrupole deformation, obtained from an energy density functional at the mean-field level, into an effective Hamiltonian expressed as a many-body operator. The advantage of this procedure is that one avoids the problems with density dependence which can arise in beyond-mean-field methods. The effective Hamiltonian is then used in a straightforward way in the generator-coordinate-method with the inclusion of projections onto good particle numbers and angular momentum. In the end, both spectra and wave functions are obtained. As an example of the method, calculations for the nucleus $ ^{62} $Zn is performed with three different parametrizations of the Skyrme functional. The results are compared with experiment.
\end{abstract}

\section{Introduction}
A successful route to understanding and describe atomic nuclei is the mean-field approach. In this view, all the nucleons build up a common potential in which they move independently. This potential can, for example, be found in a self-consistent way through the Hartree-Fock (HF) method or, with the inclusion of pairing correlations, the Hartree-Fock-Bogoliubov (HFB) method. In those methods, the state of the nucleus is one single Slater-determinant of single-particle or quasiparticle nature.

The mean-field approach is able to describe bulk properties of nuclei, such as the binding energy, with high accuracy. Also, it is common that the solutions break one or more symmetries of the underlying interaction. The most prominent is the breaking of spherical symmetry into a deformed nucleus. This symmetry breaking provides a way to gain physical insight into nuclear phenomena. For example, by interpreting the deformation as a static shape of the nucleus in its intrinsic frame, one can understand the emergence of rotational bands which show up in many nuclear spectra.

However, despite the success of the mean-field description, it can not be the complete description of the system. Due to the breaking of symmetries, it is impossible for the solutions to be true eigenstates of the nuclear Hamiltonian. One would expect a true eigenstate, which respects the symmetries, to be a super position of different shapes with different orientations. Furthermore, a deformed state can not be labeled with good quantum numbers for the angular momentum. And that is crucial in order to get proper theoretical spectra and to calculate transitions between states; without the use of extra assumptions for going from the intrinsic frame into the laboratory frame. 

The method presented here belongs to the class of beyond-mean-field methods. The common feature for those methods is that the state of the nucleus can be expressed as a linear combination of several Slater-determinants; as opposite to the single one in mean-field methods.

In our model, we want to restore the broken symmetries of particle numbers and angular momentum with the use of projections. Also, we want to include correlations associated with large amplitude collective motions; that is: vibrations and rotations. This is done through the generator-coordinate-method (GCM) together with the cranking method. As the generator coordinates we use the $ \beta $ and $ \gamma $ deformation parameters.

In the search of a theoretical method to describe and predict low energy properties of atomic nuclei, a common starting point is to introduce a general two- and three-body interaction. In order to perform practical computations, the three-body interaction is often converted to a density dependent two-body interaction; turning the interaction into an energy density functional (EDF)~\cite{Bender2003}. A commonly used EDF is based on the Skyrme expansion~\cite{SKYRME}.

Using an EDF in the framework of HFB will reproduce experimental binding energies with high precisions. Therefore, it can be a good idea to use those results as input to a beyond-mean-field method.

However, extending the mean-field models based on EDF:s with projections and GCM has been shown to give rise to formal and technical issues. Firstly, some effective interactions, which are used for the low-momentum regime of nuclei, do also contain a non-physical high-momentum part. This part is not probed at the mean field level; whereas it can be so in GCM when mixing different many-body states. Those interactions can give rise to ultraviolet divergences~\cite{Carlsson2013}.

Secondly, how to treat the density in the mixing process is not well defined; although many approximate ways have been proposed~\cite{Satula,Duguet,Waroquier,Bender2009}.

Therefore it can be important to have the interaction expressed as a many-body Hamiltonian with no density dependence when aiming for beyond-mean-field calculations.

\section{Model}
In our model, we use results from an EDF as input to a beyond-mean-field calculation through the frameworks of GCM and projections. In the end, the model produces spectra and wave functions. The wave functions can be used to calculate observables; such as electromagnetic transitions. In the future, we want to use the wave functions as input to reaction studies by constructing an optical potential out of them; see Ref.~\cite{Idini} in this volume.

A detailed description of the model can be found in Ref.~\cite{Ljungberg}. Here follows a summary.

\subsection{The Effective Hamiltonian}
The starting point for this model is to postulate an effective Hamiltonian of the form
\begin{equation}
H_{eff}=\sum_{i}e_{i}a^{\dagger}_{i}a_{i}-
\\
\frac{1}{4}G\sum_{ijkl}P_{ij}P_{kl}a^{\dagger}_{i}a^{\dagger}_{j}a_{l}a_{k}-
\\
\frac{1}{4}\chi\sum_{\mu}\sum_{ijkl}\left[ \tilde{Q}^{2\mu}_{ik}\tilde{Q}^{2\mu*}_{lj}-\tilde{Q}^{2\mu}_{il}\tilde{Q}^{2\mu*}_{kj} \right]a^{\dagger}_{i}a^{\dagger}_{j}a_{l}a_{k}.
\end{equation}
Here, the first term is taken from a spherical HF calculation based on an EDF. This calculation constructs a single-particle basis for which $ a^{\dagger}_{i} $ is a creation operator for the spherical orbital $ i $ with energy $ e_{i} $. The second term is a pairing term. We use the seniority pairing interaction with the strength $ G $; which is determined from the uniform spectra method~\cite{nilsson1995}. The third term is a modified quadrupole interaction adopted from Ref. \cite{KUMAR1970}. At the mean-field level, this interaction gives rise to a deformed Wood-Saxon potential. The strength of the quadrupole interaction, $ \chi $, is a free parameter. It will be determined through a fit to constrained EDF calculations.

The form of $ H_{eff} $ has been chosen in such a way that it includes the most important features of atomic nuclei. Namely: spherical single-particle energies, pairing correlations and quadrupole deformations. 

\subsection{Mapping an EDF into $ H_{eff} $}
The parameter $ \chi $ is selected such that $ H_{eff} $ reproduces results obtained from an EDF. That is: Mapping an EDF into $ H_{eff} $. Here we choose to calculate the binding energy as a function of the quadrupole deformation, $ E(\beta_{2}) $, from constrained HF calculations; both with an EDF and with $ H_{eff} $. Then $ \chi $ is selected such that the two curves for $ E(\beta_{2}) $ coincides as much as possible.

\subsection{GCM and Projections}
\label{H-W}
With $ \chi $ fixed, $ H_{eff} $ can be used in the framework of GCM in a straightforward way. In GCM, the eigenenergies and eigenstates are extracted from the space spanned by a many-body basis. In our construction of this basis we include cranking. We define a cranked Hamiltonian as $ H_{eff}^{\omega}\equiv H_{eff}-\omega I_{x} $ with some constraint on $ \expval{I_{x}} $. Here $ I_{x} $ is the angular momentum around the $ x $-axis and $ \omega $ is the angular frequency which acts like a Lagrange multiplier. The basis states are obtained by solving the HFB equations for $ H_{eff}^{\omega} $ with different constraints on the $ \beta $ and $ \gamma $ deformation parameters; generating a many-body basis of HFB vacua spread out over the $ (\beta,\gamma) $-plane. In addition, each HFB vacuum is assigned different values for $ \expval{I_{x}} $ and for the pairing gaps, which are used to determine $ G $, of both neutrons and protons. Those values are randomly drawn from a set of a few allowed values for each parameter.

In the framework of GCM, the matrix elements of $ H_{eff} $ and the overlap, $ \mathcal{O} $, in the many-body basis $ \lbrace\ket{\phi_{a}}\rbrace $ are needed. For the overlap we use the new formula of Ref.~\cite{Overlap} in our model.
 
At this stage, projections onto good particle numbers and angular momentum is performed. Hence, the matrix elements to be calculated are
\begin{equation}
H_{a'K',aK}=\mel{\phi_{a'}}{H_{eff}P^{I}_{K'K}P^{N}P^{Z}}{\phi_{a}}, 
\\
 \hspace{10mm} \mathcal{O}_{a'K',aK}=\mel{\phi_{a'}}{P^{I}_{K'K}P^{N}P^{Z}}{\phi_{a}}.
\end{equation}
Here $ P^{I}_{K'K} $ is the projection operator for total angular momentum $ I $ with projection $ K (K') $ on the  symmetry-axis for state $ \ket{\phi_{a}} (\ket{\phi_{a'}}) $; and $ P^{N/Z} $ is the projection operator for particle number of neutrons/protons. See Ref.~\cite{Bally} for the explicit expressions of those projection operators and how to discretize them in numerical computations.

In the end, the spectrum and wave functions, for a given angular momentum, are obtained from the Hill-Wheeler equation
\begin{equation}
\sum_{aK}H_{a'K',aK}h_{aK}^{n}=E_{n}\sum_{aK}\mathcal{O}_{a'K',aK}h_{aK}^{n}
\end{equation}
where $ E_{n} $ is the energy and $ \lbrace h_{aK}^{n}\rbrace $ are the expansion coefficients for for the $ n $:th state in the basis of projected HFB states. Hence, the final state can be written
\begin{equation}
\ket{IM,n}=\sum_{aK}h_{aK}^{n}P^{I}_{MK}P^{N}P^{Z}\ket{\phi_{a}}
\end{equation}
where $ M $ is the projection of the angular momentum on the laboratory $ z $-axis.

In practice, the Hill-Wheeler equation is solved by first diagonalizing the overlap matrix to obtain the so called natural basis. Then, in that basis, the Hamiltonian is diagonalized. However, this can be numerically unstable because of the overcompleteness of the many-body basis. Therefore we diagonalize the Hamiltonian repeatedly. First in a small basis of just a few natural states that have the largest norms. Then adding one state for each diagonalization; sorted by decreasing norm. A converged result is indicated by the formation of a plateau; where the addition of states no longer change the energy.

\section{Results}
As a demonstration of our method, and to give examples of the outputs from the model, calculations have been performed for the nucleus $ ^{62}\text{Zn} $. For this nucleus it exists a rich amount of data for which the theoretical results can be compared to. The calculations has been done with three different parametrizations of the Skyrme functional: UNE1, SLy4 and SKM*.

\subsection{Mapping}
\begin{figure}
\begin{center}
\includegraphics[scale=0.335]{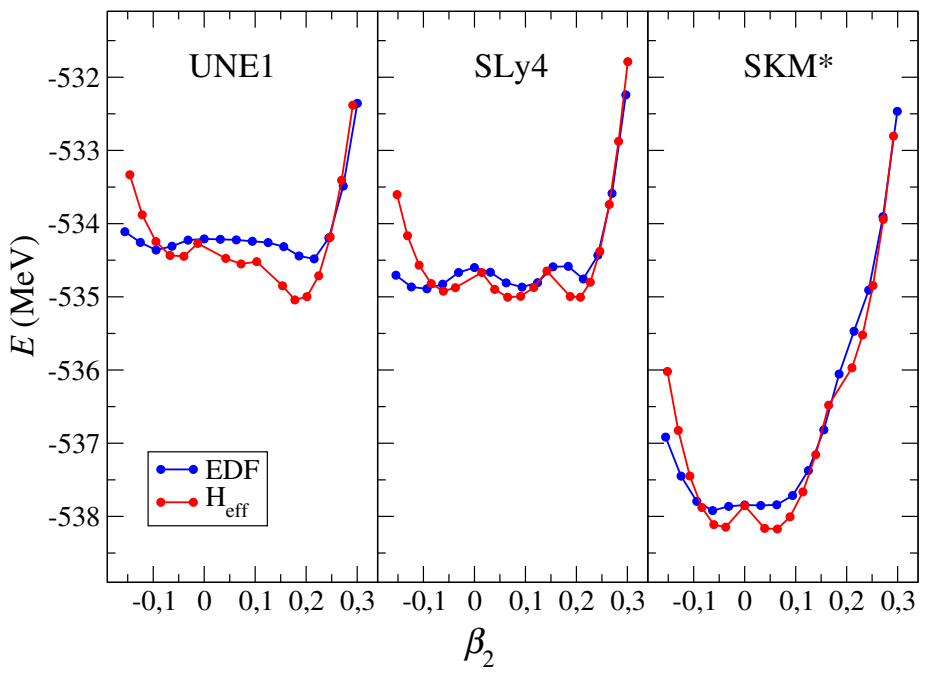}
\caption{Fits of the parameter $ \chi $ for the three parametrizations of the Skyrme functional given at the top of each plot. The blue curves are the HF energies obtained from the EDF; whereas the red curves are HF energies obtained from $ H_{eff} $. The energy is the binding energy; which can be compared to the experimental value:~$ E=-538.1 $ MeV \cite{Experiment}. Since pairing correlations are not included, we expect the HF energies to be higher than the experimental one.\label{fig:fit}}
\end{center}
\end{figure}

For the HF calculations, 12 major shells in the spherical oscillator basis has been used. The EDF results are from the code HFBTHO \cite{HFBTHO}; whereas the results from $ H_{eff} $ are obtained from an updated version of the code HOSPHE \cite{hosphe}. The curve $ E(\beta_{2}) $ has been fitted between $ -0.15\leq\beta_{2}\leq0.30 $. In Fig.~\ref{fig:fit} the fits for the three parametrizations are presented.
Even though the three parametrizations generate quite different $ E(\beta_{2}) $ curves, $ H_{eff} $ can capture  them all in a qualitative way.

\subsection{The Many-Body Basis}
For UNE1, the $ (\beta,\gamma) $-plane was sampled with 250 points within $ -30^{\circ}\leq\gamma\leq150^{\circ} $  and $ \beta\leq0.36 $. Only states below the excitation energy $ E^{*}=16 $ MeV were kept. This resulted in basis size of 209 states.

For SLy4 and SKM*, the $ (\beta,\gamma) $-plane was sampled with 400 points within $ -30^{\circ}\leq\gamma\leq150^{\circ} $ and $ \beta\leq $(0.60~/~0.48) for (SLy4~/~SKM*). Furthermore, a quadratic cut in the excitation energy was chosen; only states below $ E^{*}=[8+0.07\times I_{x}(I_{x}+1)] $ MeV were kept. The number 0.07 was selected to fit with the experimental yrast band. This resulted in a basis size of (151~/~145) states for (SLy4~/~SKM*).

For the discretization of the projection onto good particle numbers, 10 points was used. For the angular momentum, 28 points for each Euler angle was used for UNE1; whereas 32 points was used for SLy4 and SKM*. All spins up to $ 10\hbar $ was projected out.

\subsection{Spectra}
The ground state of $ ^{62} $Zn can be viewed as a core plus six valence nucleons, four neutrons and two protons, in the $ fp $-group above the 28 gap. At low excitation energies, the valence nucleons are limited to this group. The maximal spin that can be built in this configuration, by aligning all valence nucleons, is $ 10\hbar $. Hence, one expects the yrast band to terminate at $ 10\hbar $. At higher energies two neutrons, to ensure positive parity, can be excited into the $ g_{9/2} $ subshell. This configuration terminates at $ 16\hbar $. In this work we focus on the low energy regime up to $ I=10\hbar $.

\begin{figure}
\begin{center}
\includegraphics[scale=0.335]{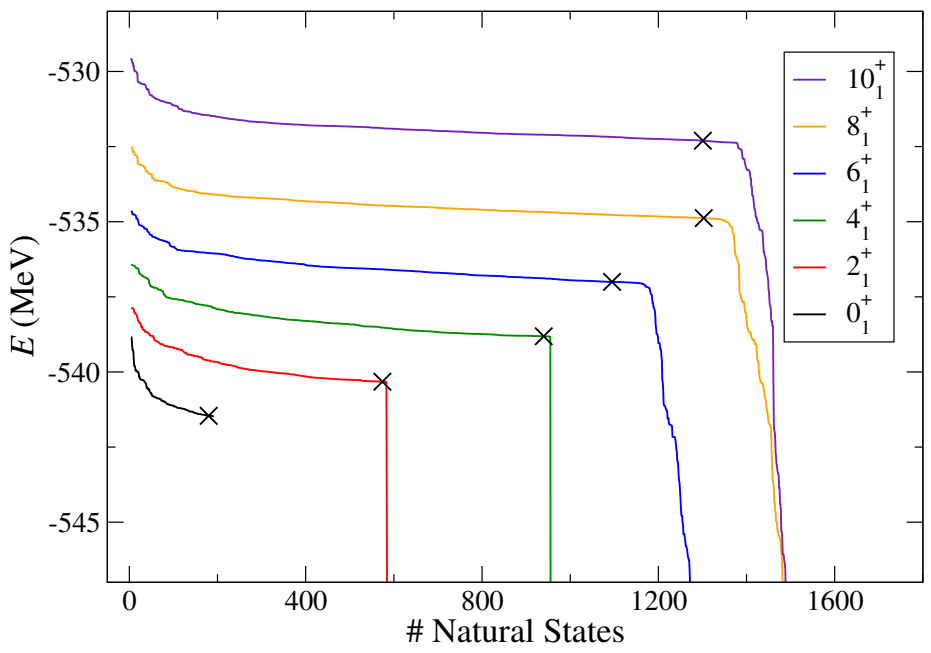}
\caption{The energy obtained from diagonalizing the Hamiltonian in a basis with different numbers of natural states. The curves are for the yrast states of even spins. The crosses show where the solutions have been selected. This is from the calculation built upon UNE1. \label{fig:platt}}
\end{center}
\end{figure}

The spectra is calculated in the way described in section~\ref{H-W}. As a visualization of the procedure, Fig.~\ref{fig:platt} shows the energy obtained, for the yrast states of even angular momentum, with different sizes of the natural basis. This is from the calculation built upon UNE1.
The region where the solution becomes unstable is distinct for all spins above zero. The solutions are taken just before those regions; shown by the crosses.

\begin{figure}[!h]
\includegraphics[width=18.8pc]{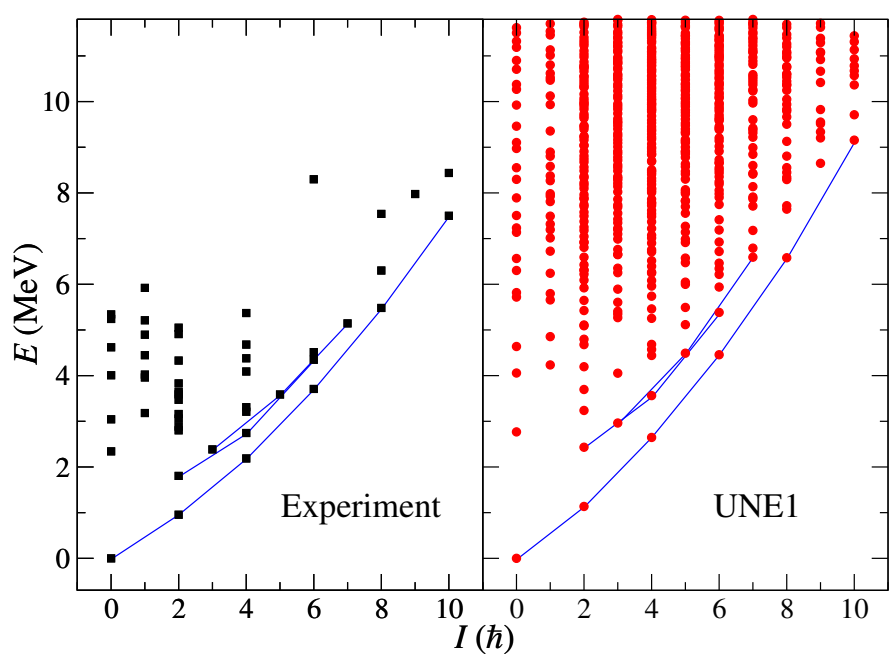}
\includegraphics[width=18.8pc]{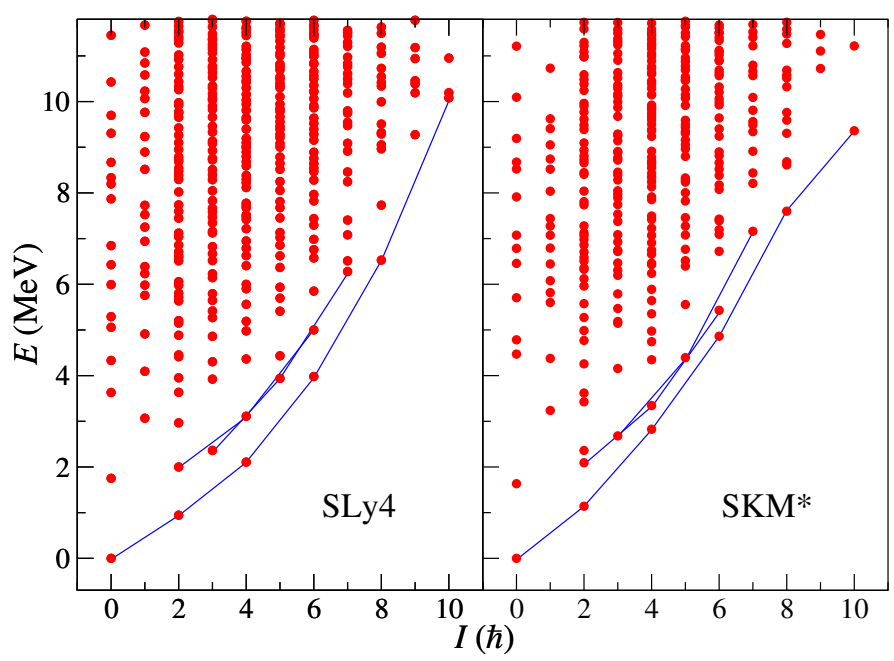}
\caption{The spectrum for $ ^{62} $Zn from experiment~\cite{Experiment} and from our model built upon the parametrizations UNE1, SLy4 and SKM*. For the experimental data, only states with positive parity and $ I\leq10\hbar $ are shown. The blue lines represent the bands which are found in experiment and their corresponding positions in the theoretical spectra.\label{fig:spectra}}
\end{figure}
The final results for the spectra is shown in Fig.~\ref{fig:spectra}. Also experimental values, taken from~\cite{Experiment}, are shown for comparison. First consider the UNE1 parametrization. The overall structure close to the yrast band is in good agreement with experiments; even though the energies increases too fast with angular momentum. It is interesting to compare the  band structure. In experiment, there is a clear yrast band up to $ I=10\hbar $ and two bands on top of each other just above the yrast one; ranging from $ I=2\hbar $ to $ I=7\hbar $. This structure of states is reproduced by the calculation. Compared to our previous work~\cite{Ljungberg}, we have included odd spin states in the model. The agreement with experiment for those states indicates that our method holds also for the odd spin case.

For the other parametrizations, SLy4 gives very similar spectra as UNE1. Hence, for those parametrizations, the method is not too sensitive to the underlying EDF it is founded on.

In contrast to that, SKM* does not succeed as well as the other parametrizations to reproduce the structure above the yrast band. And the increase in energy with angular momentum is even higher; in accordance with the different energy landscape in Fig.~\ref{fig:fit}.

\section{Discussion}
A beyond-mean-field method has been described. This method uses the results from an EDF to constrain an effective Hamiltonian. This Hamiltonian is then used in the framework of GCM, in order to include correlations associated with vibrations and rotations, and projections are used to restore symmetries. The form of the effective Hamiltonian allows for many-body calculations in a straightforward way.

The results shown indicate, despite the simplicity of the interaction, that the effective Hamiltonian capture enough physics in order to reproduce experimental data. And also, because of this simplicity, the model can be a valuable tool for the description of heavy and superheavy nuclei~\cite{dirk}.

\section*{References}
\bibliography{iopart-num}

\end{document}